\documentclass[%
 reprint,
 amsmath,amssymb,
 aip
]{revtex4-2}

\usepackage{graphicx}
\usepackage[utf8]{inputenc}

\begin{document}
\author{M.\ Bia\l{}ek}
\email{marcin.bialek@fuw.edu.pl}
\address{Institute of Physics, \'Ecole Polytechnique F\'ed\'erale de Lausanne (EPFL), 1015 Lausanne, Switzerland}
\author{J.\ Zhang}
\address{Fert Beijing Institute, MIIT Key Laboratory of Spintronics, School of Integrated Circuit Science and Engineering, Beihang University, Beijing, China}
\author{H.\ Yu}
\address{Fert Beijing Institute, MIIT Key Laboratory of Spintronics, School of Integrated Circuit Science and Engineering, Beihang University, Beijing, China}
\address{International Quantum Academy, Shenzhen 518048, China}
\author{J.-Ph.\ Ansermet}
\address{Institute of Physics, \'Ecole Polytechnique F\'ed\'erale de Lausanne (EPFL), 1015 Lausanne, Switzerland}
\address{International Quantum Academy, Shenzhen 518048, China}

\title{Antiferromagnetic resonance in $\alpha$-Fe$_2$O$_3$ up to its N\'eel temperature}

\date{\today}

\begin{abstract} 
Hematite ($\alpha$-Fe$_2$O$_3$) is an antiferromagnetic material with a very low spin damping and high N\'eel temperature. The temperature dependence of the antiferromagnetic resonance in a bulk single crystal of hematite was characterized from room temperature up to the N\'eel temperature in the frequency range of 0.19--0.5~THz. From these data, the N\'eel temperature was estimated as 966~K.
\end{abstract}

\maketitle

%\section{Introduction}

Hematite ($\alpha$-Fe$_2$O$_{3}$) is a very common room-temperature antiferromagnet of the N\'eel temperature, reported as $T_N=946$~K,\cite{Oravova13} $T_N=950$~K,\cite{Aleksandrov85} $953$~K,\cite{Morin50} $960$~K,\cite{Neskovic77} $972$~K\cite{Ono62}. It is characterized by a very low spin damping that make it promising for spintronic applications \cite{Sulymenko17, Stremoukhov19, Fischer20, Lebrun20, Grishunin21, Ross21} and strong light-mater coupling.\cite{Bialek21, Boventer22} Above the spin-reorientation transition (Morin phase transition) at about $T_M=260$~K,\cite{Aleksandrov85, Chou12} the superexchange Dzialoshinskii-Moriya interaction leads to canting of the two sublattices that gives rise to net magnetisation $\mathbf{m}$, i.e. making this material a weak ferromagnet.

Hematite crystallizes in an approximately hexagonal structure with space group R$\overline{3}$c. Precise measurements show that the actual symmetry is monoclinic C2/c \cite{Przenioslo14}.
In the weak ferromagnetic state the ferromagnetic moment $\mathbf{m}$ is along $a$ axis\cite{DZYALOSHINSKY58, Morrish} (magnetic symmetry C2'/c' \cite{Przenioslo14}) or $b$ axis\cite{DZYALOSHINSKY58, Aleksandrov85} (magnetic symmetry C2/c)\cite{Oravova13, Przenioslo14}). Owing to the spin canting, the antiferromagnetic resonance has two modes: at a higher frequency, the quasi-antiferromagnetic resonance (qAFMR) mode which is excited when the dynamical magnetic field $\mathbf{h}$ is parallel to the magnetization ($\mathbf{h}\parallel \mathbf{m}$), and at low frequencies, the quasiferromagnetic resonance (qFMR) mode, excited by a dynamical magnetic field perpendicular to the magnetization ($\mathbf{h}\perp \mathbf{m}$). In this work, we characterize the qAFMR mode up to the N\'eel temperature, which was observed previously up to about 600~K\cite{Aleksandrov85} and below room temperature\cite{Chou12}.

The qFMR mode has non-zero frequency only with external field applied\cite{Velikov69, Zavislyak19} and falls below our experimental range. To our knowledge, the qFMR mode was investigated only at room temperature and below\cite{Morrison73, Boventer21, Wang21}. %With regard to the spin dynamics, hematite can be considered as a low loss material , resulting in very narrow width of the antiferromagnetic resonance. %more and more %neutron scattering 25km/s

%\section{Experimental}
Development of frequency extenders for vector network analyzers (VNA) allows continuous-wave spectroscopic measurements up to 1.5~THz to be conducted with a high frequency resolution and with a very high dynamic range \cite{Caspers16, Bialek18, Bialek19, Bialek20, Bialek21}. Our sample was a natural single crystal of $\alpha$-Fe$_2$O$_{3}$ of $d=0.5$~mm in thickness and 10$\times$10~mm$^2$ in lateral dimensions.
The normal to the sample surface was (11-20). 

%When measuring transmission through bulk samples, the crystal was placed in-between the two oversized cylindrical metallic waveguides of 11~mm inner diameter. These were fixed inside a cylindrical ceramic furnace under PID-controlled temperature, measured with a K-type thermocouple positioned close to the sample.

\begin{figure}
\begin{center}
\includegraphics[width=\linewidth]{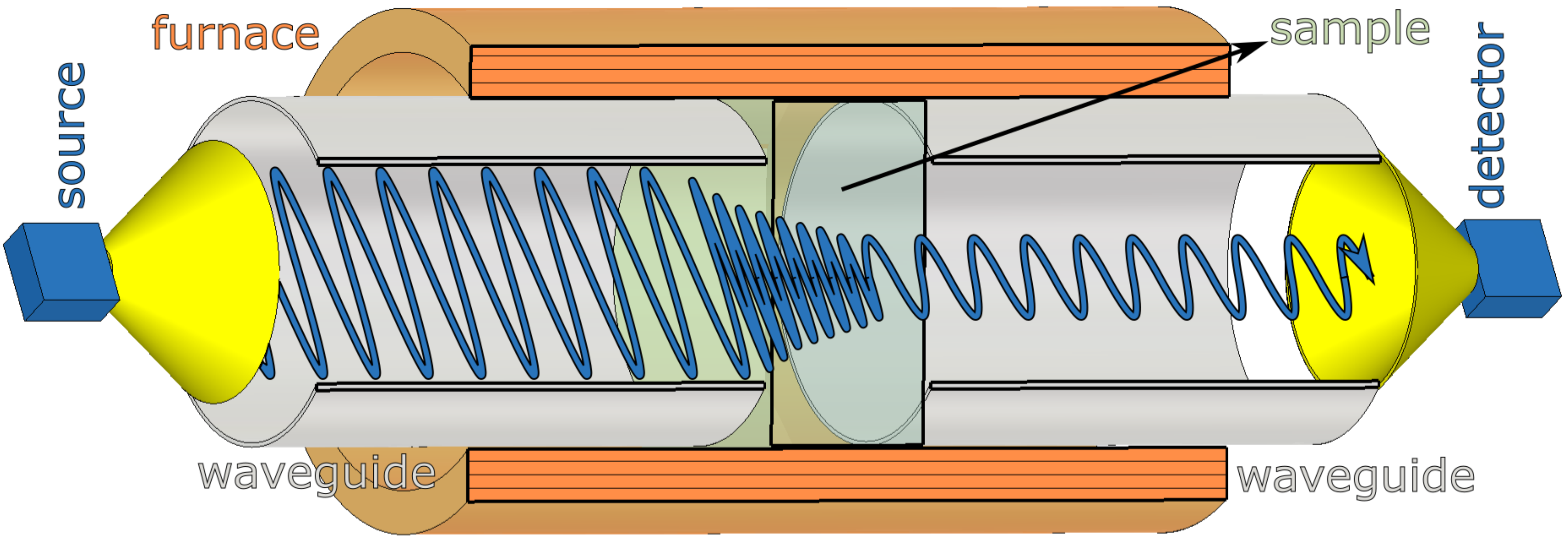}
\caption{{\label{setups}
Schematic of the THz spectrometer (not to scale): source and detector VNA extenders, horns, oversized waveguides, the sample is heated with a furnace.
}}
\end{center}
\end{figure}
Figure \ref{setups} shows a schematic of our THz setup, where terahertz radiation generated by the source extender was guided to the sample using oversized metallic waveguides. Transmitted radiation was guided from the sample to the detector using identical oversized metallic waveguides. Detector measured incoming power and phase of the electric field. Temperature scans started from the highest temperature with a step of $\Delta T=-0.25$~K. We chose $\Delta T$ so that change of the resonance frequency with temperature is smaller than the line width of the resonance, which is particularly narrow at lower temperatures. After stabilizing sample temperature $T$ with a PID controller, we measured transmission as a function of radiation frequency $f$. The recorded spectra were obtained by averaging 20 frequency sweeps measured with 100 Hz bandwidth of extenders intermediate frequency bandwidth. %that is just a technical VNA detail
This procedure was applied at two frequency bands of 0.19--0.35~THz and 0.33--0.5~THz using different sets of frequency-extenders. We recorded relative  transmitted radiation power in $dB$ and the phase of the transmitted electric field in $deg$ units. This way, a complex transmission data matrix $S_{21}(f,T)$ was obtained.

The source extender emits linearly polarized beam of THz radiation and the detector extender detects only radiation of a linear polarization that matches its waveguide. Rotation of the polarization plane about the optical axis is possible by rotating the source and detector about the optical axis. The data presented here, were obtained with parallel source and detector polarizations. %We found that the oversized waveguides that we used do not mix linear polarization significantly to our measurements.

%\section{Results}
\begin{figure*}
\begin{center}
\includegraphics[width=\linewidth]{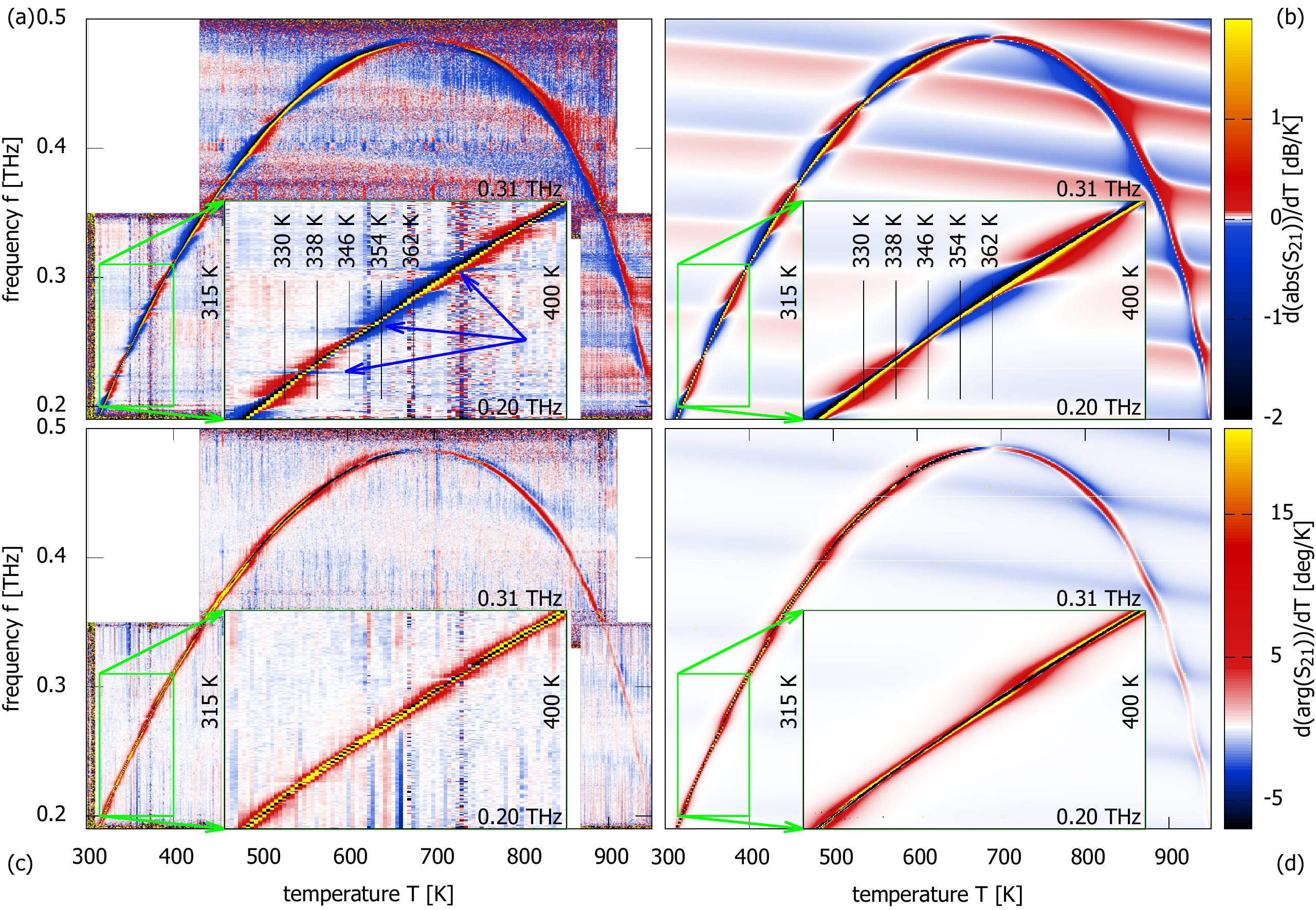}
\caption{{\label{res-bulk}
 Temperature-differential transmission through a bulk crystal, magnitude (a) and phase (c) data. Fit of the global model: magnitude (b) and phase (d)using Eq.~\ref{differential}--\ref{g_T} with parameters from Tab.~\ref{params}. Insets : zoom on a narrow temperature window, temperature marks in (a) and (b) are those of the line plots in Fig.\ \ref{params-plot}. Blue arrows point to stronger interaction with cavity modes of the setup.
}}
\end{center}
\end{figure*}
Raw spectra are dominated by interference occurring between components of the experimental setup. Since this pattern very weakly depends on temperature, we eliminated it by calculating temperature-derivative spectra, that is, by subtracting successive spectra from one another. We calculated magnitude temperature-derivative spectra
\begin{equation}
  \frac{d |S_{21}|}{d T} = \frac{|S_{21}(f,T+\Delta T)|-|S_{21}(f,T)|}{\Delta T},
  \label{dS21dTmag}
\end{equation}
and phase derivative:
\begin{equation}
  \frac{d(\arg{S_{21}})}{d T} = \frac{arg(S_{21}(f,T+\Delta T))-arg(S_{21}(f,T)))}{\Delta T}.
  \label{dS21dTpha}
\end{equation}

\begin{figure*}
\begin{center}
\includegraphics[width=\linewidth]{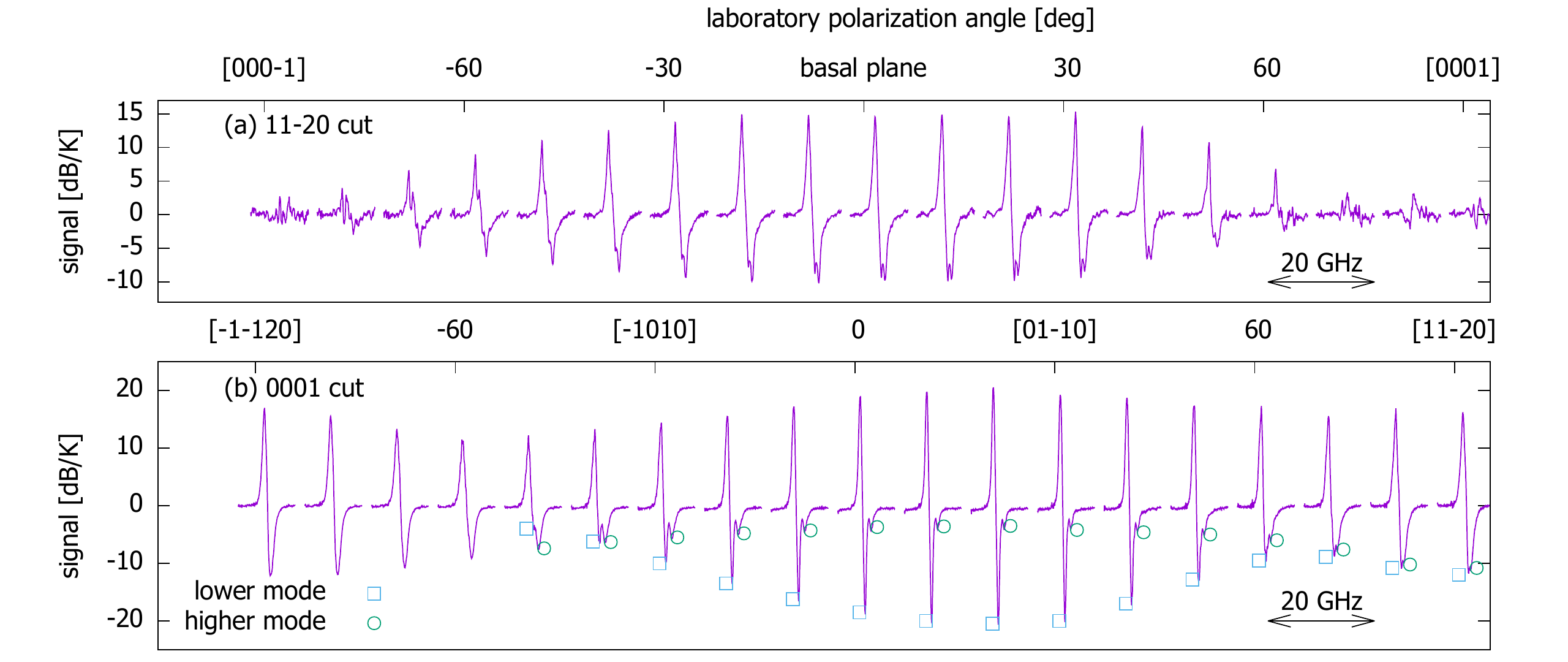}
\caption{{\label{angular}
Temperature-differential spectra near the resonance measured at different polarization angles for (a) 11-20 cut sample at 324.0~K and for (b) 0001 cut sample at 321.5~K. Spectra are shifted for clarity to show dependence of the resonance magnitude on polarization angle. In (a), for $\alpha=0^{\circ}$, THz magnetic field is in the basal plane and for $|\alpha|=90^{\circ}$ it is along $c$ axis. In (b), the points are guide for the eye, highlighting the amplitudes of qAFMR lower and higher modes.
}}
\end{center}
\end{figure*}
Our main results are presented in Fig.\ \ref{res-bulk}a (magnitude) and Fig.\ \ref{res-bulk}c (phase). They were obtained with the THz magnetic field acting in the basal plane of the crystal, as determined from polarization-dependent results in Fig.\ \ref{angular}.
At a perpendicular polarization, with THz magnetic field along $c$ axis, the resonance is almost not excited. We can notice the the resonance peak shows a tiny splitting of about 1 GHz.
%Our angular result obtained with a crystal of another cut confirms that the qAFMR is excited by dynamical magnetic field oriented within the basal plane.
Result of angular dependence obtained with a $c$-cut sample shows a more complex dependence on polarization angle that is different for a higher and a lower modes of the qAFMR (green circles and blue rectangles in Fig.\ \ref{angular}b). This angular dependence suggests that the observed splitting is related to different magnetic domains in our samples. 
%This means that our samples ar of C2/c symmetry with the weak ferromagnetic moment $m$ in the basal plane

%\section{Analysis}
We observed a strong and narrow resonance (Fig.~\ref{res-bulk}b and c), the frequency of which is rising with temperature at above room temperature  (Fig.~\ref{res-bulk}a), reaching a maximum of about 484~GHz at about 685~K and then dropping sharply as $T_N$ is approached. The periodic, almost horizontal pattern is related to the Fabry-P\'erot type cavity modes inside the sample. We can account for these temperature-differential spectra, as shown in Fig.~\ref{res-bulk} (b, d), by using an electrodynamics-based model \cite{Bialek20}:
\begin{equation}
\begin{aligned}
  {\frac{d|S_{21}(f,T)|}{d T}}= \frac{20}{\Delta T} log_{10}\frac{|t(f,T+\Delta T)|}{|t(f,T)|},
\end{aligned}
\label{differential}
\end{equation}
where $t$ is the transmittance of a plane electromagnetic wave at normal-incident on a parallel-plane slab of infinite lateral dimensions and thickness $d$
\begin{equation}
t = \frac{(1-r^2)e^{ikd}}{1-r^2e^{2ikd}}.
\end{equation}
Here, $r=(\sqrt{\epsilon}-\sqrt{\mu})/(\sqrt{\epsilon}+\sqrt{\mu})$ and $k=2\pi f\sqrt{\epsilon\mu}/c$. We assume an isotropic slab material characterized by its permeability $\mu(f,T)$ and permittivity $\epsilon(f,T)$. Our crystal is not isotropic, but in the case of incidence at a normal angle to the crystal surface, the crystal being cut in one of its principal axis, we approximate it as isotropic, that is, we assume that there is no rotation of polarization in the crystal. At other angles of incident beam polarization, we get different values of $\epsilon$. However, as long as we were interested in the behaviour of the antiferromagnetic resonance, we determined permittivity at the polarization angle where the resonance is excited the strongest. Since this occures when the dynamical magnetic field is in the basal plane, the $\epsilon$ that we determined in this communication describes response of hematite when exposed to dynamical electric field acting in the $c$ axis.

The assumptions made for both $\mu$ and $\epsilon$ are the following. We found that our data are sensitive to a temperature dependence of the dielectric response function. The simplest model that accounts for our data is a real second order polynomial of temperature. This approximation is justified because our frequency range is far from optical phonons (the lowest at about 7.5~THz) \cite{Faria97, Chamritski05, Jubb10} and dielectric absorption is negligible. Thus, we assumed
\begin{equation}
    \epsilon = \epsilon_{900}+a(T-T_0)+b(T-T_0)^2,
    \label{eps}
\end{equation}
where $T_0 = 900$~K, $\epsilon_{900}=23.405$, $a=1.139\cdot 10^{-2}K^{-1}$, $b=6.434\cdot10^{-6}$K$^{-2}$. 
To account for the magnetic resonance, we assumed that the permittivity has the form of two Lorentzian functions to account for the observed splitting of the qAFMR mode
\begin{equation}
    \mu = 1 + \frac{\Delta\mu_1 f_r^2}{f_r^2-f^2-ifg_1} + \frac{\Delta\mu_2 (f_r+\Delta f)^2}{(f_r+\Delta f)^2-f^2-ifg_2}.
\end{equation}
Here, $\Delta\mu_1 + \Delta\mu_2=\Delta\mu$ gives the total static magnetic susceptibility, $f_r$ is the lower qAFMR frequency, $g_1$ and $g_2$ are widths of the two split resonances and $\Delta f$ is the splitting of about 1~GHz. In the next paragraphs we describe these parameters.

\begin{figure}
\begin{center}
\includegraphics[width=\columnwidth]{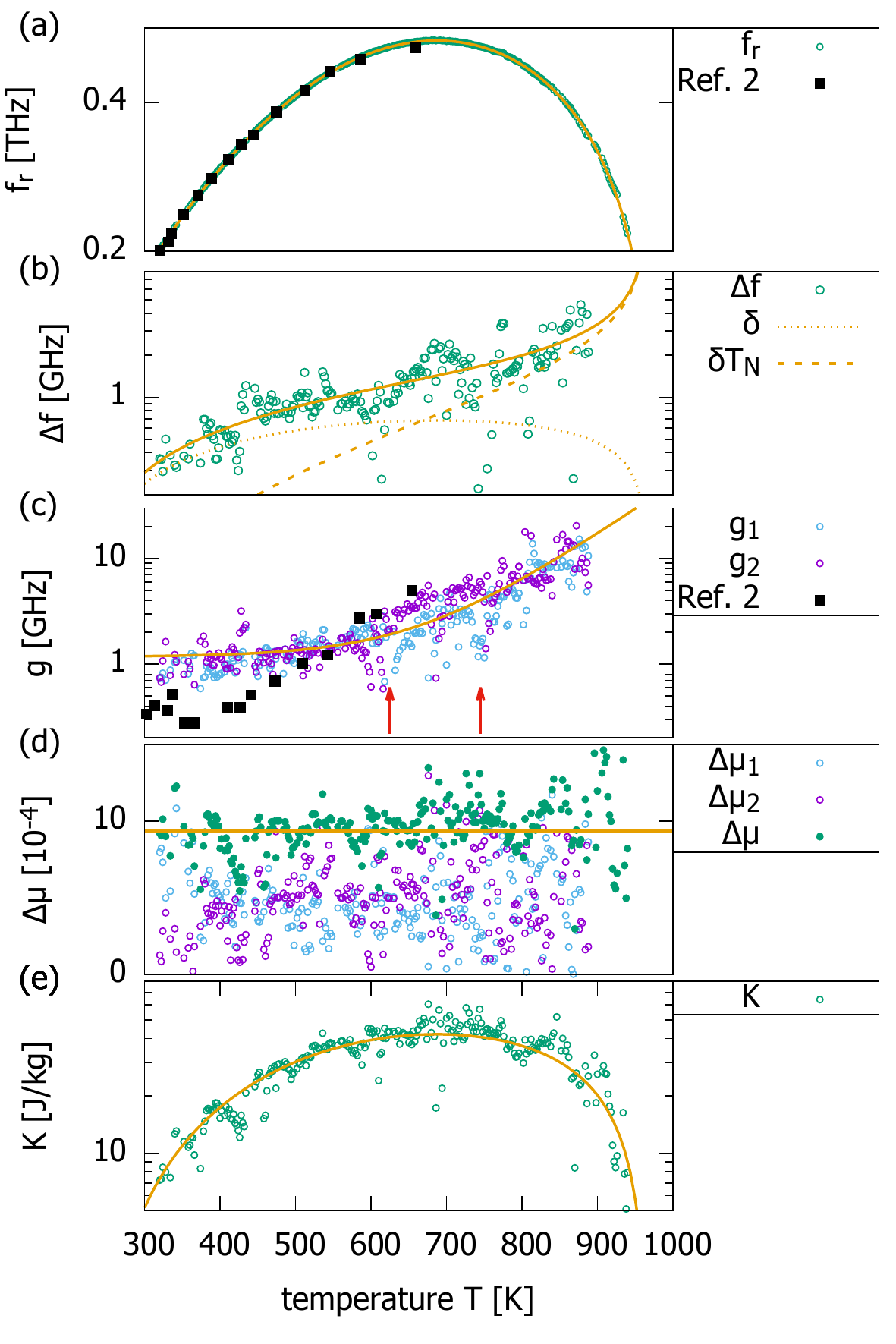}
\caption{{\label{params-plot}
Temperature dependence of the resonance frequency $f_r$ (a), amplitude $\Delta\mu_1$ and $\Delta\mu_2$ (b), width $g_1$ and $g_2$ (c), frequency splitting $\Delta f$ (d), and magnetic anisotropy factor $K$ (d). Points come from 2~K intervals fits. The solid lines are calculated using Eq.~\ref{f_T}--\ref{Kittel} with parameters from Tab.~\ref{params}. In (a) and (c), black rectangles come from Ref.~\onlinecite{Aleksandrov85}. 
}}
\end{center}
\end{figure}

In order to describe the temperature dependence of the antiferromagnetic resonance, we first fitted the resonance frequency, amplitude, and line widths in temperature intervals of 2~K around some temperature $T_0^{(l)}$, where $l$ is the interval number. Within each such interval, we put amplitudes and widths independent of temperature and we fit the resonance frequency with a first order polynomial $f(T)^{(l)}=f_{T0}^{(l)}+f_T^{(l)}(T-T0)$. Results of these temperature interval fittings are shown as green crosses in Fig.~\ref{res-bulk}(a,b,c). We show in Fig.~\ref{params-plot}(a) middle frequency $f_{T0}^{(l)}$ obtained in given temperature interval.
\begin{table}
\centering
%\begin{center}
\begin{tabular}{ ccc } 
 \hline\hline
 $a_2$      & $-3.2\pm0.3$  &          GHz$^2$K$^{-2}$\\
 $a_3$      & $1.836\pm0.017$ & 10$^{-2}$GHz$^2$K$^{-3}$\\
 $a_4$      & $-2.40\pm0.03$  & 10$^{-5}$GHz$^2$K$^{-4}$\\
 $a_5$      & $1.113\pm0.019$ & 10$^{-8}$GHz$^2$K$^{-5}$\\
 $\beta$    & $0.499\pm0.005$ & \\
 $T_N$      & $965.9\pm0.4$   & K\\
 $\Delta\mu$& $9.4\pm0.24$   & $10^{-4}$\\ 
 $g_{400}$  & $1.17\pm0.15$   & GHz\\
 $g_e$      & $34\pm3$        & GHz\\
 $g_s$      & $1.12\pm0.07$   & $10^{-2}$K$^{-1}$\\
 $\delta$   & $1.41\pm0.11$   & $10^{-3}$\\
$\delta T_N$& $1.21\pm0.06$   & K\\
 \hline\hline
\end{tabular}
%\end{center}
\caption{\label{params}Obtained global fit parameters.}
\end{table}

%\section{Global model}
These results allowed us to assume some simple functional dependences that are approximately valid in the entire experimental temperature range. 
The global fitting result, using the above assumptions, are displayed in Fig.\ref{res-bulk}(b) and (d). The fit parameter values are given in Tab.~\ref{params}. The functions used to describe modes frequencies, amplitudes and widths are drawn as solid yellow lines in Fig.~\ref{params-plot}.

%\subsection{Frequency}
To describe the observed temperature dependence of the resonant frequency (Fig.\ \ref{params-plot}a), we took the dependence of the resonant frequency to follow a modified power law \cite{Eibschutz66}
\begin{equation}
    f_r(T) = \sqrt{\sum_{j=2}^{j=5}a_jT^j}\left(1-\frac{T}{T_N}\right)^{\beta},
    \label{f_T}
\end{equation}
where $T_N$ is a parameter close to the N\'eel temperature and the power factor $\beta$ is expected to be about $1/2$. We modified the classical power law adding a 5-th order polynomial that lets us heuristically describe its nontrivial dependence on temperature.
This non-monotonic dependence of resonance frequency on temperature  is caused by the strong temperature dependence of the magnetic anisotropy field\cite{Ovchinnikov10} that governs the frequency of this mode. It is related to the spin-reorientation transition at about 260~K \cite{Aleksandrov85}, with one of its effects being that the qAFMR mode frequency has a minimum at this temperature. 
The fit value of $T_N\approx966$~K is within the spread of literature values of $T_N = (946-972)$~K\cite{Oravova13, Morin50, Neskovic77, Ono62}. 
These inconsistencies in the literature might stem from errors in temperature measurement in different experiments. However, it may be also due to different models used to determine $T_N$, when it is determined by extrapolation. 

%\subsection{Splitting}
We assumed that the observed splitting (Fig.\ \ref{params-plot}b) can be described in the following way
\begin{equation}
    \Delta f = \delta f_r + f_r(T)_{T_N=T_N+\delta T_N/2} - f_r(T)_{T_N=T_N-\delta T_N/2},
    \label{df}
\end{equation}
which express our expectation that the splitting is caused by presence of two types of domains characterized by slightly different anisotropy fields (factor $\delta$) and slightly different critical temperatures (factor $\delta T_N$). The second term dominates at high temperatures. Inhomogeneous nature of this splitting is justified by the measurements for 0001 cut sample (Fig. \ref{angular}b) that show that upper and lower modes have different dependence on polarization angle. The observed temperature dependence of the splitting can be well fitted (Fig.~\ref{params-plot}b) with the Eq.\ \ref{df} using reasonable values of $\delta$ and $\delta T_N$ (Tab.~\ref{params}). 

%\subsection{Amplitude}
Taking into account the scatter in Fig.\ \ref{params-plot}d, we took amplitudes of the resonance $\Delta\mu_1=\Delta\mu_2=\frac{1}{2}\Delta\mu$ to be independent of temperature. This scatter is probably due to imperfect interval fits, especially pronounced at around 700~K and close to $T_N$, owing to low signal obtained by temperature-differential transmission in these temperature ranges. %It is worth to notice that in Fig.\ \ref{params-plot}b the scatter of a sum of independent fit parameters $\Delta\mu_1+\Delta\mu_2=\Delta\mu$ is smaller than the scatter of each of these two parameters.

%\subsection{Width}
We assumed that the temperature dependence of the observed width (Fig.\ \ref{params-plot}c) of both modes is described by the same function consisting of a sum of a constant and an exponential,
\begin{equation}
    g(T) = g_{400} + g_ee^{g_s(T-T_{N})},
    \label{g_T}
\end{equation}
where the latter dominates close to $T_N$. It could be due to either: 1) temperature inhomogenity, 2) inhomogenity of $T_N$, 3) homogeneous line broadening. We can exclude the 1st possibility, as if it was the dominant mechanism, the linewidth would be proportional to $|df_r/dT|$, and we would see a drastic drop in the linewidth around 700~K, where $df_r/dT=0$. For the 2nd possibility we can estimate the distribution of $T_N$ that could cause such a broadening by solving $f_r(T)\pm \frac{1}{2}g_ee^{g_s(T-T_N)}=\sqrt{\sum_{j=2}^{j=5}a_jT^j}(1-T/T_{N\pm})^{\beta}$. The maximum $T_{N+}-T_{N-}$ that is necessary to account for the observed broadening is about $6$~K, which is much larger than estimation of $\delta T_N=1.2$~K obtained from the splitting of the resonance. Therefore, we conclude that the observed broadening with temperature is mainly due to rising spin damping.
%Finally, we cannot exclude that the observed broadening is due to magnetization dynamics near the phase transition, since it is difficult to measure an actual N\'eel temperature distribution in the sample, using for example magnetometry.

We notice that the linewidth drops strongly around 620~K and 740~K (two arrows in Fig.\ \ref{params-plot}c). Since at these two temperatures, the frequency of the resonance is about 475~GHz, we can guess that this effect is due to interaction of the magnetic resonance with either some electromagnetic mode in the setup or some electric transition in the crystal that has a temperature-independent frequency. %We cannot identify the nature of the cavity that is causing these narrowings. %Unfortunatelly, we don't have means to investigate further 

\begin{figure}
\begin{center}
\includegraphics[width=\columnwidth]{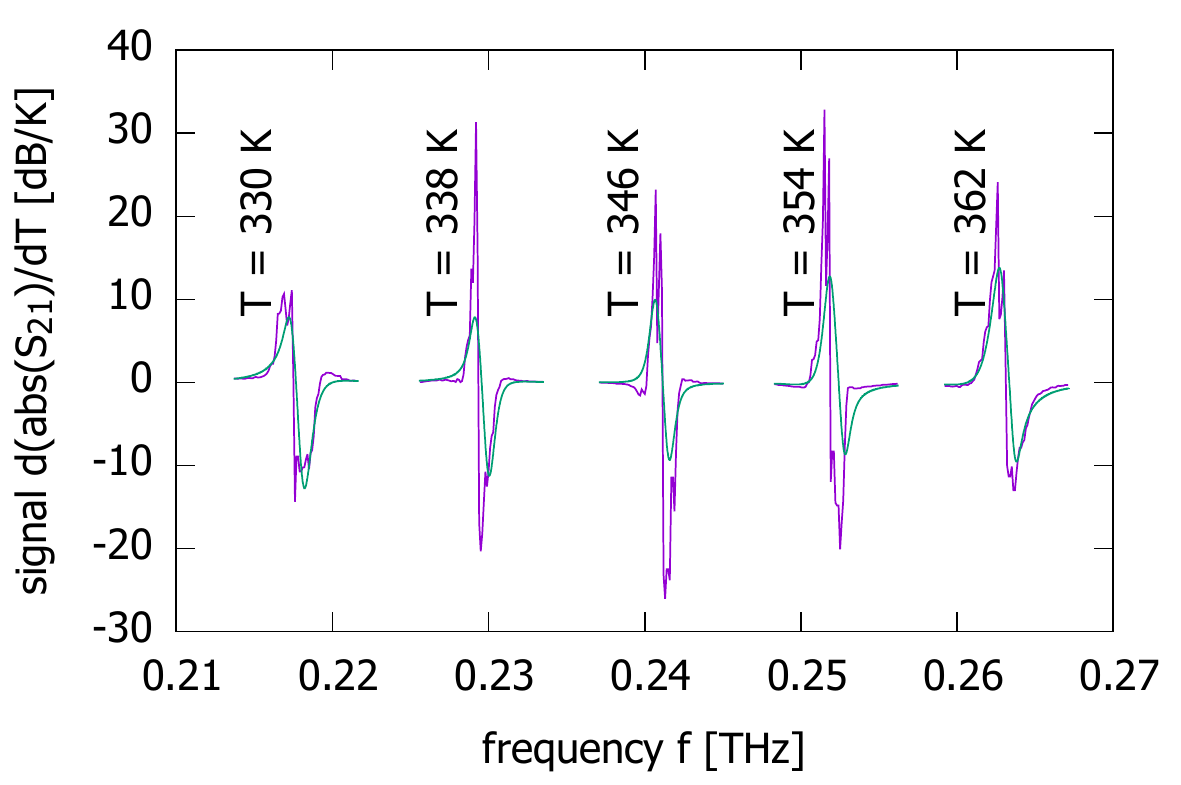}
\caption{{\label{selected-spectra}
Experimental spectra (purple) and the global fitting result (green) at selected temperatures, marked in insets of Fig.\ \ref{res-bulk}ab.  
}}
\end{center}
\end{figure}

%\section{Discussion}
There are features in Fig.~\ref{res-bulk} that are possibly caused by interaction with cavity modes than cannot be taken into account within our one-dimensional model. The effects of some of such modes are shown with blue arrows in the inset of Fig.\ \ref{res-bulk}a and a particular example of a lineshape distortion is shown in Fig.\ \ref{selected-spectra} at $T=346$ and $354$~K.
%In Fig.~\ref{selected-spectra}, we show measured spectra at different temperatures to illustrate the precision of our analysis.
We can clearly see that, at different temperatures, the lineshape changes and some of these features are not reproduced by our global one-dimensional model. Blue arrows in Fig.\ \ref{res-bulk}a show that these are indeed related to modes of cavities that cannot be taken into account by the global model, which show-up as almost temperature-indepedent horizontal lines distorting the resonance in the inset of Fig.\ \ref{res-bulk}a that are not reproduced by the fit in inset of Fig.\ \ref{res-bulk}b. These distortions cause apparent fluctuations of width and amplitude of the resonance line in the obtained temperature-interval fit values.

%From Fig.\ \ref{selected-spectra} we can also observed that the global fit does not reproduce the observed splitting perfectly. It is rather that it is a factor that allows to better describe the lineshape.

We observed weak interaction of magnetic resonance with electromagnetic standing waves, as found previously \cite{Bialek20}. These effects are readily taken into account by a classical electrodynamic model. Light-matter interaction when the sample is its own cavity is in a weak regime, because the quality factor of modes of dielectric cavity is low. Nevertheless, these electromagnetic cavity modes already have a dramatic and nontrivial effect on the observed lineshapes, as shown in Fig.~\ref{params-plot}b and Fig.~\ref{params-plot}c.

In Fig.~\ref{params-plot}e we also included the magnetic anisotropy factor, which we deduced from equation $K=\mu_0H_AM_0$, where $H_A$ is anisotropy field and $M_0$ is a saturation magnetisation of a sublattice. To do so, we took into account that the static magnetic susceptibility is given by $\chi_{\perp}=M_0/2H_E$, where $H_E$ is exchange field, and that $\chi_{\perp}=\Delta\mu_2/\rho_m$, where $\rho_m=5.27\cdot10^3$~kg/m$^3$ is the hematite mass density \cite{Pailhe08}. Then, using the Kittel's equation \begin{equation}
    2\pi f_r=\gamma\mu_0\sqrt{2H_EH_A}=\gamma\sqrt{\frac{K\mu_0}{\chi_{\perp}}},
    \label{Kittel}
\end{equation}
we found the mass density of the magnetic anisotropy energy. We see in Fig.~\ref{params-plot}e that $K$ has a complex behaviour, which is known to be caused by the spin-reorientation transition that occurs at about 260~K.

%\section{Summary}
Hematite single crystals present a very narrow resonance around room temperature, about 1~GHz. The resonance frequency increases sharply above room temperature and reaches a maximum of 484 GHz at about 685~K and then, the mode softens steeply when approaching the N\'eel temperature $T_N$. Above 690~K, the observed width of the resonance rises exponentially with temperature. Weak light-matter coupling was observed between the qAFMR mode and the standing waves inside the slab itself. The qAFMR mode presents a splitting that was ascribed to domains in the sample.

%\section*{Acknowledgements}
Support by the Sino-Swiss Science and Technology Cooperation (SSSTC) grant no.\ EG-CN\_02\_032019 is gratefully acknowledged. The VNA and frequency extenders were funded by EPFL and the SNF R'Equip under Grant No.\ 206021\_144983.

%\section*{Author declatations}
The authors have no conflicts to disclose. The data that support the findings of this study are available from the corresponding author upon reasonable request.

\bibliographystyle{unsrt}
\bibliography{refs}

\end{document}